\documentclass[a4paper,12pt]{article}
\pdfoutput=1
\usepackage{amssymb}
\usepackage{epsfig}
\usepackage{graphicx}

\makeatletter

\@addtoreset{equation}{section}
\makeatother
\def\be{\begin{equation}}
\def\ee{\end{equation}}
\def\bea{\begin{eqnarray}}
\def\eea{\end{eqnarray}}

\def\({\left(}
\def\){\right)}
\def\<{\left<}
\def\>{\right>}

\def\tr{{\mbox{tr}}}
\def\be{\begin{equation}}
\def\ee{\end{equation}}
\def\bea{\begin{eqnarray*}}
\def\eea{\end{eqnarray*}}
\def\ben{\begin{eqnarray}}
\def\een{\end{eqnarray}}
\def\({\left(}
\def\){\right)}
\def\<{\left<}
\def\>{\right>}
\def\!{\right|}
\def\|{\left|}

\def\[{\left[}
\def\]{\right]}

\def\+{\bar}
\def\mb{\mathbb}
\def\tr{{\mbox{tr}}}

\def\L{{\cal{L}}}
\def\t{\widetilde}

\def\L{{\cal{L}}}

\def\eps{{\cal{\varepsilon}}}

\begin{document}

\setlength{\unitlength}{1mm}

\pagestyle{empty}
\vskip-10pt
\vskip-10pt
\hfill %{\tt hep-th/yymmnnn}
\begin{center}
\vskip 3truecm
{\Large \bf
A nonabelian M5 brane Lagrangian
\vskip0.5truecm
in a supergravity background}
\vskip 2truecm
{\large \bf
Andreas Gustavsson}\vspace{1cm} 
\begin{center} 
Physics Department, University of Seoul, 13 Siripdae, Seoul 130-743 Korea
\end{center}
\vskip 0.7truecm
\begin{center}
(\tt agbrev@gmail.com)
\end{center}
\end{center}
\vskip 2truecm
{\abstract We present a nonabelian Lagrangian that appears to have $(2,0)$ superconformal symmetry and that can be coupled to a supergravity background. But for our construction to work, we have to break this superconformal symmetry by imposing as a constraint on top of the Lagrangian that the fields have vanishing Lie derivatives along a Killing direction.}

\vfill
\vskip4pt
\eject
\pagestyle{plain}

\section{Introduction}
Finding the nonabelian M5 brane Lagrangian is a long-standing problem, but at the same time it has also been clear for a long time that a unique classial nonabelian Lagrangian for a selfdual tensor field with manifest $(2,0)$ superconformal symmetry can not exist \cite{Witten:1996hc}, \cite{Lambert:2019diy} and we will review the argument below. With the discovery of the M2 brane Lagrangians \cite{Bagger:2007jr}, \cite{Gustavsson:2007vu}, \cite{Aharony:2008ug} a new hope was that also the M5 brane Lagrangian may be found if one relaxes some of the symmetries that should be present in the classical Lagrangian in the same spirit as one did for the ABJM Lagrangian \cite{Aharony:2008ug} of multiple M2 branes that preserves only a subgroup of the $SO(8)$ R-symmetry group. The worldvolume theory of flat M2's has the bosonic symmetry group of $AdS_4 \times S^7$. Since $S^7$ is a Hopf fiber bundle over $\mb{CP}^2$ there is a way of breaking its isometry group $SO(8)$ down to $SU(4)\times U(1)$ corresponding to this Hopf fibration and it is only this latter R-symmetry that is manifest in the ABJM Lagrangian. For the M5's on the other hand, we have the bosonic symmetry group of $AdS_7 \times S^4$ but here $S^4$ is not a Hopf circle-bundle so for the M5's it may be better to consider an orbifolding of the $AdS_7$ space which reduces the Lorentz symmetry rather than the R-symmetry. We will not attempt to orbifold $AdS_7$ in this paper, but we will consider a nonabelian theory that breaks the Lorentz symmetry at the classical level of the Lagrangian. More generally we will present a candidate Lagrangian for M5's on Lorentzian six-manifolds that has at least one Killing vector field that corresponds to an isometry direction. We break translational symmetry along this isometry direction by keeping only the zero modes of the fields in this direction. This isometry direction can be quite general. It can be fibered over a five-manifold. It can be a compact circle direction or it can be a noncompact direction. It can be of any signature, timelike, spacelike or null. The only thing that we demand is that all fields have vanishing Lie derivates along this isometry direction. This approach to the M5's has been studied previously \cite{Lambert:2010wm}, \cite{Lambert:2019diy}, \cite{Gustavsson:2018rcc} but in this paper we generalize these results and obtain a nonabelian Lagrangian coupled to supergravity background fields. This is a generalization of the abelian M5 brane coupled to a supergravity background fields that was studied in \cite{Bergshoeff:1999db}.

If we put Lie derivatives to zero in one spatial direction, then the theory will become five-dimensional and this should be nothing but 5d SYM coupled to supergravity fields \cite{Cordova:2013bea}, although expressed in a six-dimensional language. However, the Killing vector field can be of any signature and in particular it can be light-like \cite{Lambert:2010wm}, \cite{Kucharski:2017jwv}. So our Lagrangian is more general than the Lagrangian of 5d SYM. But at the quantum level the distinction between 5d SYM and the M5 brane is blurred since we do not understand whether these could in fact be just different faces of one and the same theory \cite{Douglas:2010iu}, \cite{Lambert:2010iw}. Our Lagrangian appears to have 6d $(2,0)$ superconformal symmetry. But this symmetry is broken at the classical level as we shall constrain the fields to have vanishing Lie derivatives in one direction. The hope is that this broken symmetry will be restored in the quantum theory and that instanton particles of 5d SYM will give us those missing momentum modes along the isometry direction. Supersymmetry variations for such a theory have been found previosly for special cases. First for flat $\mb{R}^{1,5}$ in \cite{Lambert:2010wm} and then for $(1,0)$ superconformal symmetry on generic circle-bundle manifolds in \cite{Gustavsson:2018rcc}. The corresponding Lagrangian for this supersymmetric system has been unknown for some time. Recently there was an interesting suggestion for such a Lagrangian in flat $\mb{R}^{1,5}$ in \cite{Lambert:2019diy} based on a construction of a Lagrangian for a selfdual tensor field that had appeared in \cite{Sen:2019qit}, \cite{Sen:2015nph}. 

In \cite{Witten:1996hc} it was argued that we shall not attempt to write down a Lagrangian for a selfdual tensor field since the partition function for a selfdual tensor field when put on an euclidean six-manifold that has three-cycles is not unique. If the partition function is not unique, then a path integral argument suggests that also the Lagrangian can also not be unique. So we shall not look for a unique Lagrangian for the selfdual tensor field. Instead we may start with quantizing a nonchiral theory and at the end perform a holomorphic factorization to select a partition function for the M5 brane theory. %There is indeed a supersymmetric abelian Lagrangian that one can write down using a nonselfdual tensor field, see for example \cite{Bak:2016vpi}. Such a nonchiral Lagrangian also appeared in \cite{Bergshoeff:1999db} where it was claimed to be not supersymmetric unless on imposes a selfduality constraint on the tensor field on top of the Lagrangian, but that appears to be a misconception since the Lagrangian is also supersymmetric without imposing such a selfduality constraint. It would be nice if that Lagrangian could be generalized to the nonabelian case. But we could not find such a supersymmetric nonabelian Lagrangian that remains supersymmetric also when we turn on the supergravity background fields, the R-gauge potential $V_M$ and the supergravity background tensor field $T^A_{MNP}$. So instead 

In this paper we will go against this philosophy, at least naively, and instead we will consider the Lagrangian for a selfdual tensor field that was found in \cite{Sen:2019qit}, \cite{Sen:2015nph}. It appears that this Lagrangian can be supersymmetrized and then it might also have applications to the M5 brane system \cite{Lambert:2019diy}, \cite{Andriolo:2020ykk}. 

The objection rised by the paper \cite{Witten:1996hc} to the study of Lagrangians for selfdual tensor fields, can be avoided when there is a Killing direction in the six-manifold that might select one partition function as special compared to the many other partition functions that may also appear. For instance, if this Killing vector is timelike, then we may put time along this Killing vector and use Hamiltonian quantization that will give us a unique partition function. The canonical example for this approach is the M5 brane on a flat six-torus where Hamiltonian quantization selects for us a unique the partition function, among several candidate partition functions, that is the one that happens to also be modular invariant \cite{Dolan:1998qk}. In fact our Lagrangian, that depends on a choice of Killing vector field, may also fit well with the idea of \cite{Witten:1996hc} after all, because from this work one is just discouraged to go looking for a unique Lagrangian for the selfdual tensor field. Our  Lagrangian is not necessarily unique. If there are several Killing vector fields then there is one Lagrangian for each choice of `preferred' Killing vector field that is used to construct our Lagrangian. This is in the same spirit as that of Hamiltonian quantization, but here generalized to Killing vectors that can be either timelike, spacelike or null leading to more general quantizations than the usual Hamiltonian quantization that applies only for the case of a timelike Killing vector.

Any proposed Lagrangian for the M5's can be put to the following tests. The first and simplest test of any candidate $(1,0)$ supersymmetric Lagrangian is whether this can be enhanced to $(2,0)$. There are several attempted nonabelian M5 brane Lagrangians in the literature that do not appear to pass this test \cite{Gustavsson:2018xjr}, \cite{Ho:2014eoa}, \cite{Samtleben:2011fj} although that does not rule out the more exotic possibilty (actually realized by ABJM theory) that supersymmetry could get enhanced to $(2,0)$ at the quantum level. Another test is whether any attempted $(2,0)$ Lagrangian in flat space can be put on curved space and whether $(2,0)$ supersymmetry can be enhanced to $(2,0)$ superconformal symmetry. Finally one may test whether a given candidate Lagrangian can be consistently coupled to the eleven-dimensional supergravity background fields while preserving superconformal symmetry. 

In this paper we will present a Lagrangian that appears to pass all these tests, but this is not entirely correct because for this construction to work we need to impose as a constraint on top of the Lagrangian that the Lie derivatives of all the fields vanish along a Killing direction and thus we need to break some of the superconformal symmetry at the classical level. But a breaking of some of the spacetime symmetries at the classical level of a Lagrangian is precisely what we should expect as that enables us to have a classical Lagrangian description of the M5's that is not unique, but depends on a choice of Killing vector.

\section{The supersymmetric Lagrangian}
Following \cite{Lambert:2019diy}, \cite{Sen:2019qit}, \cite{Sen:2015nph} we introduce a selfdual tensor field $H^+_{MNP}$. This is an auxiliary tensor three-form field whose role in the Lagrangian is as a Lagrange multiplier field that implements the selfduality condition on another three-form field that we will denote as $g_{MNP}$. Part of $g_{MNP}$ is a three-form $h_{MNP}$ with the wrong sign kinetic term in the Lagrangian. For abelian gauge group this three-form is a field strength of a two-form gauge potential $b_{MN}$, so that $h_{MNP} = 3\nabla_{[M} b_{NP]}$. For the nonabelian generalization we will not present an explicit realization of $h_{MNP}$ in terms of some nonabelian two-form gauge potential. This is one of the longstanding mysterious aspects of the theory of multiple M5 branes, the mystery of what exactly would be the nonabelian two-form. We will not try to answer this question here. But we will postulate the the infinitesimal variation can be presented as
\ben
\delta h_{MNP} &=& 3 D_{[M} \delta b_{NP]}\label{deltab}
\een
for some infinitesimal nonabelian two-form variation $\delta b_{MN}$. Let us assume that $h_{MNP} v^P = - F_{MN}$. Let us define the gauge algebra valued one-form
\ben
Y^T &:=& \eps^{MNPRST} [h_{MNP},F_{RS}]\label{hF}
\een
Dualizing (\ref{hF}) we get
\bea
[h_{[MNP},F_{RS]}] &=& - \frac{1}{120} \eps_{MNPRST} Y^T
\eea
Moreover $v^M [h_{[MNP},F_{RS]}] = 0$ since $v^M F_{MN} = \L_v A_N = 0$ and we assume that $h_{MNP} v^P = - F_{MN}$. This is realized by taking $Y^T \sim v^T$. Conversely, if $Y^T$ has another component not parallel to $v^T$ then we get $\eps_{MNPRST} Y^T v^M \neq 0$. So we have
\bea
Y^M &=& v^M Y
\eea
for some gauge algebra valued zero-form $Y$. We now get
\bea
D_M (v^M Y) = v^M D_M Y = \L_v Y = 0
\eea
once we impose the gauge fixing condition $v^M A_M = 0$. We get zero because we constrain the Lie derivative along $v^M$ of all fields to vanish, so in particular $\L_v Y = 0$. Using the Bianchi identity $D_{[T} F_{RS]} = 0$ we now get 
\ben
\eps^{MNPRST} [F_{RS},D_{T} h_{MNP}] &=& 0\label{FDh}
\een
This does does not necessarily imply that $D_{[T} h_{MNP]} = 0$ since to derive (\ref{FDh}) we have assumed that $F_{RS} = - h_{RST} v^T$, so $F_{RS}$ can not be varied independently from $h_{MNP}$. If $v^2 \neq 0$ then we have the projection operators 
\bea
P_M^N &=& \delta_M^N - \frac{1}{v^2} v_M v^N\cr
Q_M^N &=& \frac{1}{v^2} v_M v^N
\eea
that enable us to decompose
\bea
h_{MNP} &=& h'_{MNP} - F_{[MN} v_{P]}
\eea
where 
\bea
h'_{MNP} &=& P_P^Q h_{MNQ}
\eea
Now $h'_{MNP}$ and $F_{MN}$ can be varied independently from each other. Also, we notice that
\bea
\eps^{MNPRST} [F_{RS},D_{T} \(F_{MN} v_P\)] = \eps^{MNPRST} [F_{RS},F_{MN}] \nabla_T v_P = 0
\eea
simply because $\eps^{MNPRST} = \eps^{PRMNST}$. Then we have left 
\bea
\eps^{MNPRST} [F_{RS},D_{T} h'_{MNP}] &=& 0
\eea
and since $F_{RS}$ is independent from $h'_{MNP}$, we conclude that
\bea
D_{[T} h'_{MNP]} &=& 0
\eea
Hence
\ben
D_{[T} h_{MNP]} &=& \frac{1}{2} F_{MN} w_{PT}\label{bhid}
\een
where we define
\bea
w_{MN} &=& \nabla_M v_N - \nabla_N v_M
\eea

To formulate the supersymmetry variations, we find it convenient to introduce an infinitesimal variation $\delta B_{MN} := - \delta b_{MN}$. When this variation is a supersymmetry variation, then this is given by $\delta B_{MN} = i \bar\eps\Gamma_{MN}\psi$. But this is just the infinitesimal variation, and we do not introduce nonabelian gauge potentials $b_{MN}$ nor $B_{MN}$ in this paper, only their infinitesimal variations.

The Lagrangian is a sum of two terms, $\L = \L_b + \L_m$ where the gauge field part is 
\bea
\L_b &=& \frac{1}{24} h_{MNP}^2 + \frac{1}{6} H^{+MNP} g^-_{MNP}\cr
&& + \frac{1}{6} h^{-MNP} w_{MNP} + \frac{1}{24} w_{MNP}^2\cr
&& + \lambda^{-MNP} G^+_{MNP}\cr
&& + \frac{1}{48} \eps^{MNPQRS} F_{MN} W_{PQRS}\cr
&& - \frac{1}{24 v^2} \eps^{MNPRST} \(A_M \nabla_N A_P - \frac{2i e}{3} A_M A_N A_P\) w_{RS} v_T
\eea
and the matter field part is 
\bea
\L_m &=& - \frac{1}{2} (D_M \phi^A)^2 + \frac{i}{2} \bar\psi \Gamma^M D_M \psi - \frac{1}{2} \mu^{AB} \phi^A \phi^B\cr
&& + \frac{e}{2} \bar\psi \Gamma_M \Gamma^A [\psi,\phi^A] v^M + \frac{e^2 v^2}{4} [\phi^A,\phi^B]^2\cr
&& + \frac{i}{8} \bar\psi \Gamma^{MNP} \Gamma^A \psi T^A_{MNP}\cr
&& + \frac{ie}{2} \eps^{ABCDE} \phi^E [\phi^A,\phi^B] V_M^{CD} v^M 
\eea
where we have defined 
\bea
g^-_{MNP} &=& h^-_{MNP} + w^-_{MNP} + 6 T^A_{MNP} \phi^A\cr
G^+_{MNP} &=& H^+_{MNP} + h^+_{MNP} + w^+_{MNP}
\eea
Here $G^+_{MNP}$ is a supersymmetry singlet and $g^-_{MNP} = 0$ is the selfduality equation of motion we get by varying the selfdual field $H^{+MNP}$ in the Lagrangian. We present the explicit form of the mass matrix $\mu^{AB}$ in equation (\ref{mu}).

The supersymmetry variations are 
\bea
\delta \phi^A &=& i \bar\eps \Gamma^A \psi\cr
\delta A_M &=& \delta B_{MN} v^N\cr
\delta H_{MNP}^+ &=& - \delta h^+_{MNP} - \delta w^+_{MNP}\cr
\delta h_{MNP} &=& - D_{[M} \delta B_{NP]}\cr
\delta B_{MN} &=& i \bar\eps \Gamma_{MN} \psi\cr
\delta W_{MNPQ} &=& - e \bar\eps \Gamma_{MNPQ} \Gamma^A [\psi,\phi^A]
\eea
We define 
\bea
w_{MNP} &=& W_{MNPQ} v^Q
\eea
that is a three-form with selfdual and antiselfdual components whose supersymmetry variations are 
\bea
\delta w_{MNP}^+ &=& \frac{e}{2} \bar\eps \Gamma_Q\Gamma_{MNP} [\psi,\phi^A] v^Q\cr
\delta w_{MNP}^- &=& - \frac{e}{2} \bar\eps \Gamma_{MNP} \Gamma_Q [\psi,\phi^A] v^Q
\eea
The supersymmetry variation of the fermions is 
\bea
\delta \psi &=& \frac{1}{12} \Gamma^{MNP} \eps H^+_{MNP} + \Gamma^M \Gamma^A \eps D_M \phi^A - 4 \Gamma^A \eta \phi^A - \frac{ie}{2} \Gamma_M \Gamma^{AB} \eps [\phi^A,\phi^B] v^M
\eea
Neither $\L_b$ nor $\L_m$ is supersymmetric by themselves, and only the sum is supersymmetric. 

The supersymmetry parameter satisfies the conformal Killing spinor equation 
\ben
D_M \eps &=& \Gamma_M \eta - \frac{1}{8} \Gamma^A \Gamma^{RST} \Gamma_M \eps T^A_{MNP}\label{CKE}
\een
It should be noted that this equation implies that $\Gamma^M D_M \eps = \eta$ since $\Gamma^M \Gamma^{RST} \Gamma_M = 0$. Here $T^A_{MNP}$ is a supergravity background tensor field, carrying in addition an R-symmetry vector index $A=1,...,5$. This tensor field is antiselfdual, since the spinors are chiral, 
\bea
\Gamma \eps &=& -\eps\cr
\Gamma \psi &=& \psi
\eea
where $\Gamma = \Gamma^{012345}$ is the 6d chirality gamma matrix. All our gamma matrices are eleven-dimensional, so in particular the gamma matrices for the Lorentz group and the R-symmetry group anticommute, $\{\Gamma_M,\Gamma_A\} = 0$. 

The theory also couples to the supergravity background R-gauge field $V_M^{AB}$ through the covariant derivatives that acts on the matter fields as
\bea
D_M \psi &=& \nabla_M \psi - i e [A_M,\psi] + \frac{1}{4} V_M^{AB} \Gamma^{AB} \psi\cr
D_M \phi^A &=& \nabla_M \phi^A - i e [A_M,\phi^A] + V_M^{AB} \phi^A
\eea
where $\nabla_M$ is the geometric covariant derivative that only involves the Christoffel symbol, and $e$ is an electric charge, which eventually will be fixed to some value of order one due to selfduality. But to determine the exact value of $e$ will require considerations that go beyond just classical supersymmetry so we will keep this as a free parameter here. All our fields transform in the adjoint representation of the gauge group. But this maybe can be made more general if one can find a nonabelian gerbe structure for our theory.

\subsection{Some comments}
The abelian tensor multiplet in flat space is described by its representations of the little group $SO(4)$ that preserves the momentum four vector of a massless particle in flat space. The abelian tensor multiplet contains particles that transform under this little group in the representations $(3,1)$ for the selfdual tensor field, $(1,1)$ for the five scalar field $\phi^A$ where $A=1,...,5$ and $(2,1)$ for the four fermions\footnote{Here we label representations of $SO(4)$ by the dimensions of the representations of $SU(2) \times SU(2)$}. No classical field theory description for the corresponding nonabelian tensor multiplet is known. One may find a classical field theory description for the nonabelian theory by first performing a dimensional reduction of the abelian M5 brane along a Killing vector field $v^M$ that may be either spacelike, timelike or lightlike and then one may find its nonabelian generalization there. In this paper we proceed in a different way though. We make the dimensional reduction implicit by keeping the 6d language. We have a unified form of the Lagrangian for any choice of Killing vector field along which we implicitly perform the dimensional reduction. We formulate the theory using a 6d language but our theory lives in a 5d subspace since we contrain all the fields to have vanishing Lie derivatives along the Killing vector field. Our Lagrangian is more general than the 5d SYM Lagrangian that was discovered in \cite{Cordova:2013bea} because our Lagrangian captures three different types of dimensional reductions associated to a Killing vector that is either spacelike, timelike or lightlike. By choosing our Killing vector field to be lightlike we should be able to make contact with the Lagrangian that was recently found in \cite{Lambert:2020scy}. 

We can perform explicit dimensional reduction of our Lagrangian and get a Lagrangian in 5d. But to get 5d SYM we also would need to perform a nonabelian dualization of a three-form field strength in 5d, which is a nontrivial operation in the nonabelian case. This dualization was done in the abelian case in \cite{Andriolo:2020ykk}. Direct dimensional reduction of our Lagrangian will give a dual formulation of 5d SYM. This formulation may be useful since that 5d Lagrangian contains a three-form gauge field $h_{mnp}$ coming from our $h_{MNP}$ upon dimensional reduction, where $m$ runs over five dimensional subspace. We may now use this $h_{mnp}$ to define a nonabelian Wilson surface in our 5d theory. This Wilson surface is very difficult to introduce in usual 5d SYM where there is only a one-form gauge potential. We may argue indirectly that we must get a Lagrangian that is dual to 5d SYM upon dimensional reduction as follows. We may start with abelian gauge group and perform the reduction where dualization is easy to perform following \cite{Andriolo:2020ykk}. In 5d we may subsequently find its nonabelian generalization \cite{Cordova:2013bea}, \cite{Lambert:2020scy} which is a nonabelian 5d SYM Lagrangian. But the nonabelian generalization should be unique. So we expect that the Lagrangian that we get upon direct dimensional reduction of our nonabelian Lagrangian will be dual to nonabelian 5d SYM Lagrangian, although showing that explicitly may be out of reach as that amounts to carrying out a nonabelian dualization. 

If one performs dimensional reduction along a timelike circle, then one will get an Euclidean 5d theory that came from a Lorentzian 6d theory. One may then ask how Wick rotation of the 6d theory can be implemented in the 5d theory. One proposal is that this amounts to Wick rotation of the hypermultiplet mass \cite{Minahan:2013jwa}, \cite{Gustavsson:2015fra}, \cite{Bak:2019lye}. We do not have a physical interpretation of what dimensional reduction along a compact timelike circle means. But it is the most rigorous way of deriving a supersymmetric Euclidean 5d theory from 6d which we may for example put on $S^5$. 

The physical interpretation of the dimensional reduction to 5d is best understood for a spacelike circle. The abelian tensor multiplet reduces to a vector multiplet. We have massless particles of the little group $SO(3)$ that transform in the representations $3$ (the vector representation of $SO(3)$) for the vector gauge field, $1$ (the trivial representation of $SO(3)$) for the five scalar fields and $2$ (the spinor representation of $SO(3)$) for the four fermion fields. Generalizing to the nonabelian case there are adjoint gauge indices attached to each field. The number of gauge indices in the adjoint representation of the gauge group for $SU(N)$ gauge group is $N^2-1$. The Kaluza-Klein modes that were truncated upon the dimensional reduction are believed to reappear as instanton particles and nothing gets lost in the dimensional reduction. Taking these instanton particles into account, we expect the number of components to grow like $N^3$, as was shown explicitly for the case of one instanton particle in \cite{Bak:2013bba}. 

In our formulation we have introduced nonabelian auxiliary fields $W_{MNPQ}$, $H^+_{MNP}$ and $\lambda^-_{MNP}$ (also in the adjoint representation) that appear in the Lagrangian without derivatives. They play the role of Lagrange multiplier fields that do not add any additional degrees of freedom. There is also a gauge field $A_M$ (in the adjoint representation) in our formulation, but this gauge field is not independent field from the tensor field $h_{MNP}$ as one can show by integrating out the auxiliary fields. We show this explicitly in section 6. One may still feel inconvenient with our Lagrangian since it is not familiar. So in section 7 we relate our Lagrangian to the more familiar nonchiral Lagrangian for the tensor field $h_{MNP}$ where the antiselfdual part of $h_{MNP}$ is decoupled \cite{Witten:1996hc}, by integrating out the auxiliary field $\lambda^-_{MNP}$ and subsequenly putting $W_{MNPQ} = 0$. By such a manipulation we find the right sign of the kinetic term for $h_{MNP}$. However, putting $W_{MNPQ}$ to zero is not a consistent truncation in the nonabelian case. This means that the nonchiral Lagrangian has no straightforward nonabelian generalization. This explains why our nonabelian Lagrangian has to be of a nonstandard form. 

If we choose the Killing vector field to be lightlike we get another kind of 5d SYM with a different and larger symmetry group. In particular there will be a surviving conformal symmetry \cite{Lambert:2019fne}. If the worldvolume is $\mb{R}^{1,5}$ and we reduce along a lightlike direction, it was shown there that we get 5d SYM with a bosonic $SU(3,1)$ conformal symmetry that is a subgroup of a corresponding superconformal symmery. That we are able to preserve such a large symmetry group of the full 6d superconformal symmetry as a manifest symmetry in a classical Lagrangian is rather interesting.

This work originated from our attempts to find a superconformal Lagrangian that could reproduce the equations of motion in \cite{Lambert:2010wm}, \cite{Gustavsson:2018rcc}
 that were obtained by closing supersymmetry variations on-shell. A superconformal Lagrangian can be constructed for abelian gauge group by supersymmetrizing the nonchiral Lagrangian in \cite{Witten:1996hc}, see for instance \cite{Gustavsson:2018xjr}, but this construction does not have any straightforward nonabelian generalization, which is due to a nontrivial equation of motion for the selfdual tensor field in the nonabelian case. This problem had already been addressed in \cite{Lambert:2019diy}. Their proposed Lagrangian was constructed by demanding that it gives all the desired equations of motion. One might then expect that such a Lagrangian could also be supersymmetric. However, the Lagrangian they presented does not appear to be gauge invariant. In the present paper we modify their proposed Lagrangian and we carefully check that it is superconformal. We also generalize to curved spacetime and to a supergravity background. By superconformal symmetry we here refer to a symmetry that our Lagrangian has when we allow the Killing vector field to transform under the conformal group \cite{Linander:2011jy}, \cite{Gustavsson:2018rcc}.

\section{The supersymmetry variation of $\L_m$}
For this computation, we make the ansatz
\ben
\L_m &=& - \frac{1}{2} (D_M\phi^A) + \frac{i}{2} \bar\psi\Gamma^M D_M \psi - \frac{1}{2} \mu^{AB} \phi^A \phi^B\cr
&& + \frac{a}{2} \bar\psi \Gamma_M\Gamma^A [\psi,\phi^A] v^M + \frac{b}{4} [\phi^A,\phi^B]^2\cr
&& + \frac{ic}{8} \bar\psi \Gamma^{MNP} \Gamma^A \psi T^A_{MNP}\label{matteransatz}
\een
and for the supersymmetry variation of $\psi$ we make the ansatz
\bea
\delta \psi &=& \frac{1}{12} \Gamma^{MNP} \eps H^+_{MNP} + \Gamma^M \Gamma^A \eps D_M \phi^A - 4 \Gamma^A \eta \phi^A - \frac{i d}{2} \Gamma_M \Gamma^{AB} \eps [\phi^A,\phi^B] v^M
\eea
while for the other fields we let those vary according to the what we stated before. Then we compute the supersymmetry by adopting the convention that we make integrations by parts in such a way that $\delta \psi$ does not appear in anyone of the terms and discard boundary terms. This will uniquely determine the variation as
\bea
\delta \L_m &=& D_M^2 \phi^A \delta \phi^A + i \bar\psi \Gamma^M D_M \delta \psi - \mu^{AB} \phi^A \delta \phi^B\cr
&& - i e [\delta A_M,\phi^A] D^M \phi^A + \frac{e}{2} \bar\psi \Gamma^M [\delta A_M,\psi]\cr
&& + a \bar\psi \Gamma^M \Gamma^A [\delta \psi,\phi^A] v_M + \frac{a}{2} \bar\psi \Gamma^M \Gamma^A [\psi,\delta \phi^A] v_M\cr
&& + b [\phi^A,\phi^B][\phi^A,\delta \phi^B]\cr
&& + \frac{i c}{4} \bar\psi \Gamma^{MNP} \Gamma^A \delta \psi T^A_{MNP}
\eea
We now pick the commutator terms from this variation and postpone the study of all the rest to later. Let us also study the cubic term in fermi-fields later. Then we will for now focus on the following terms in the variation of the matter fields Lagrangian
\bea
(\delta \L_m)_{comm} &=& \frac{1}{2} \bar\psi \Gamma^A \Gamma^{MN} \eps [e F_{MN} - a H_{MNP}^+ v^P,\phi^A]\cr
&& + (4a-2d) \bar\psi \Gamma^{AB} \Gamma_M \eta [\phi^A,\phi^B] v^M\cr
&& - \frac{d}{8} \bar\psi \Gamma^{AB}\Gamma^C \Gamma^{MNP} \Gamma^Q \eps [\phi^A,\phi^B] T^C_{MNP} v_Q\cr
&& - \frac{c d}{8} \bar\psi \Gamma^C\Gamma^{AB} \Gamma^{MNP} \Gamma^Q \eps [\phi^A,\phi^B] T^C_{MNP} v_Q\cr
&& + i \(a d v^2 - b\) \bar\eps \Gamma^B \psi [\phi^A,[\phi^A,\phi^B]]\cr
&& + \frac{d}{2} \bar\psi \Gamma^{MN} \Gamma^{AB} \eps [\phi^A,\phi^B] \nabla_M v_N
\eea
In addition to these terms, we also get the terms
\bea
&& d \bar\psi\Gamma^M \Gamma^N \Gamma^{AB} \eps [D_M \phi^A,\phi^B] v_N\cr
&& + e \bar\psi \Gamma^M\Gamma^N \Gamma^A \Gamma^B \eps [D_N\phi^B,\phi^A] v_M
\eea
and another such commutator term comes from 
\bea
- (\delta D_M) \phi^A D^M \phi^A &=& - e \bar\psi \Gamma_{MN} \eps [D^M \phi^A,\phi^A] v^N
\eea
so the sum of all these terms for $a=d=e$ just becomes a couple of Lie derivatives,
\bea
2 e \bar\psi \Gamma^{AB} \eps [\L_v \phi^A,\phi^B] - e \bar\psi\eps [\L_v \phi^A,\phi^A]
\eea
So we can now conclude that we shall pick 
\bea
a &=& e\cr
d &=& e\cr
b &=& e^2 v^2
\eea
To determine the value of $c$ requires some more work. This comes about by putting $4a-2d = 2e$ and then by looking at the term
\bea
2e \bar\psi \Gamma^{AB} \Gamma_M \eta [\phi^A,\phi^B] v^M
\eea
and by using the Killing spinor equation to extract from this term the following term
\bea
2 e \bar\psi \Gamma^{AB} \(v^M D_M \eps + \frac{d}{4e} \Gamma^{MN} \nabla_M v_N\) [\phi^A,\phi^B] \cr
+ \frac{e}{4} \bar\psi \Gamma^{AB}\Gamma^C \Gamma^{MNP} \Gamma^Q \eps [\phi^A,\phi^B] T^C_{MNP} v_Q
\eea
Let us make the following ansatz for the Lie derivative of a spinor field,
\bea
\L_v \chi &=& v^P D_P\chi + \frac{\alpha}{4} \Gamma^{PQ} \chi \nabla_M v_N\cr
\L_v \bar\psi &=& v^P D_P\bar\psi - \frac{\alpha}{4} \bar\psi \Gamma^{PQ} \nabla_P v_Q
\eea
Then for the vector field $T_M = \bar\psi \Gamma_M \chi$ we get
\bea
\L_v T_M &=& v^P D_P T_M + \frac{\alpha}{2} T^Q \(\nabla_M v_Q - \nabla_Q v_M\)
\eea
For this to agree with the Lie derivative of a vector field we shall have 
\bea
\alpha &=& 1\cr
\nabla_M v_N + \nabla_N v_M &=& 0
\eea
so we must now require that $v_M$ is a Killing vector field. Since the Lie derivative that we want here is 
\bea
\L_v \eps &=& v^M D_M \eps + \frac{1}{4} \Gamma^{MN} \eps \nabla_M v_N
\eea
we clearly see that we shall choose $d = e$. This second term that got generated through the usage of the Killing spinor equation now combines with the two other terms to give us 
\bea
&& - \frac{e}{8} \bar\psi \Gamma^{AB}\Gamma^C \Gamma^{MNP} \Gamma^Q \eps [\phi^A,\phi^B] T^C_{MNP} v_Q\cr
&& - \frac{c e}{8} \bar\psi \Gamma^C\Gamma^{AB} \Gamma^{MNP} \Gamma^Q \eps [\phi^A,\phi^B] T^C_{MNP} v_Q\cr
&& + \frac{e}{4} \bar\psi \Gamma^{AB}\Gamma^C \Gamma^{MNP} \Gamma^Q \eps [\phi^A,\phi^B] T^C_{MNP} v_Q
\eea
Thus for $c=1$, we get the commutator $[\Gamma^{AB},\Gamma^C] = - 4 \Gamma^{[A} \delta^{B]C}$ and the three terms collapse to 
\bea
\frac{e}{2} \bar\psi \Gamma^A \Gamma^{MNP} \Gamma^Q \eps [\phi^A,\phi^B] T^B_{MNP} v_Q
\eea
We now use the symplectic Majorana properties to write 
\bea
\bar\psi \Gamma^B \Gamma^{MNP} \Gamma^Q \eps &=& \bar\eps \Gamma^Q \Gamma^{MNP} \Gamma^B \psi 
\eea
and then we recall that 
\bea
\delta w_{MNP}^+ &=& \frac{e}{2} \bar\eps \Gamma^Q \Gamma^{MNP} \Gamma^A [\psi,\phi^A] v_Q
\eea
to write this result as 
\bea
&& \frac{e}{2} \bar\eps \Gamma^Q \Gamma^{MNP} \Gamma^A \psi [\phi^A,\phi^B] T^B_{MNP} v_Q\cr
&=& \frac{e}{2} \bar\eps \Gamma^Q \Gamma^{MNP} \Gamma^A [\psi,\phi^A]\phi^B T^B_{MNP} v_Q\cr
&=& \delta w^{+MNP} T^B_{MNP} \phi^B
\eea
We now recall that
\bea
\delta H^+_{MNP} &=& - \delta h^+_{MNP} - \delta w^+_{MNP}
\eea
By considering the abelian type of terms below, we will discover that the above variation combines with those abelian terms into 
\bea
\delta \L_m &=& - \delta \(H^{+MNP} T^B_{MNP} \phi^B\) + ...
\eea
The above computation shall be modified when there is nonvanishing background R-gauge field $V^{AB}_M$. The Lie derivatives shall be replaced as follows
\bea
\L_v \phi^A &\rightarrow & \L_v \phi^A + V^{AB} \phi^B\cr
\L_v \psi &\rightarrow & \L_v \psi + \frac{1}{4} V^{AB} \Gamma^{AB} \psi
\eea
where $V^{AB} := v^M V_M^{AB}$. Then as we put the Lie derivatives on the right side of these arrows to zero, there will survive terms  proportional to the R-gauge field in the variation of the Lagrangian. These terms are
\bea
(\delta \L_m)_V &=& \frac{e}{2} \bar\psi \Gamma^{AB} \Gamma^{CD} \eps [\phi^A,\phi^B] V^{CD} + 2 e \bar\psi \Gamma^{AB} \eps [\phi^C,\phi^B] V^{AC} + e \bar\psi \eps [\phi^A,\phi^B] V^{AB}
\eea
Using the identity
\bea
\Gamma^{AB} \Gamma^{CD} &=& - 2 \delta^{AB}_{CD} + 4 \delta^{BC} \Gamma^{AD} + \Gamma^{ABCD}
\eea
we get
\bea
(\delta \L_m)_V &=& \frac{e}{2} \bar\psi \Gamma^{ABCD} \eps [\phi^A,\phi^B] V^{CD}
\eea
that we cancel by adding the following term 
\bea
\L_V &=& \frac{i e}{6} \eps^{ABCDE} \phi^E [\phi^A,\phi^B] V^{CD}
\eea
to our ansatz for the matter fields Lagrangian in (\ref{matteransatz}), where we define $\eps^{ABCDE} := \Gamma \Gamma^{ABCDE}$.

Let us now write down the cubic terms in fermi-fields,
\bea
(\delta\L_m)_{cubic} &=& \frac{i e}{2} \bar\psi \Gamma^M [\bar\eps\Gamma_{MN}\psi v^N,\psi] + \frac{ia}{2} \bar\psi \Gamma^M\Gamma^A [\psi,\bar\eps\Gamma^A\psi]v_M
\eea
This is identically zero for $a=e$ by a Fierz identity that we derive in the appendix.

We now turn to the abelian terms, by which we refer to as those terms that will survive also when we put all the commutators to zero. Abelian terms arise from the following terms in $\delta \L_m$,
\bea
(A1) &=& D^2 \phi^A \delta \phi^A\cr
(A2) &=& i \bar\psi \Gamma^M D_M \delta \psi\cr
(A3) &=& - \mu^{AB} \phi^A \delta \phi^B\cr
(A7) &=& \frac{i}{4} \bar\psi\Gamma^{MNP}\Gamma^A\delta \psi T^A_{MNP}
\eea
We now extract all the abelian terms that will appear in each of these terms,
\bea
(A1) &=& D^2 \phi^A i \bar\eps\Gamma^A \psi
\eea
\bea
(A2a) &=& - \frac{i}{8\cdot 12} \bar\psi\Gamma^M\Gamma^{RST}\Gamma^{UVW}\Gamma_M\Gamma^A\eps T^A_{UVW} H^+_{RST}\cr
(A2b) &=& - \frac{i}{2} \bar\eps\Gamma^{NP} \psi D^M H^+_{MNP}\cr
(A2c) &=& - 4 i \bar\psi \Gamma^A \Gamma^M \eta D_M \phi^A\cr
(A2d) &=& - \frac{i}{4} \bar\psi \Gamma^{RST}\Gamma^M \Gamma^A \Gamma^B \eps T^B_{RST} D_M \phi^A\cr
(A2e) &=& - i \bar\eps \Gamma^A \psi D^2 \phi^A\cr
(A2f) &=& \frac{e}{2} \bar\psi \Gamma^A \Gamma^{MN} \eps [F_{MN},\phi^A]\cr
(A2g) &=& \frac{i}{2} \bar\psi\Gamma^A \Gamma^{MN} \eps W_{MN}^{AB}\phi^B\cr
(A2h) &=& 4 i \bar\psi \Gamma^A \Gamma^M \eta D_M \phi^A\cr
(A2i) &=& 4 i \bar\psi \Gamma^A (\Gamma^M D_M \eta) \phi^A
\eea
\bea
(A7a) &=& - \frac{i}{4\cdot 12} \bar\psi\Gamma^{UVW}\Gamma^{RST}\Gamma^A\eps T^A_{UVW} H^+_{RST}\cr
(A7b) &=& - \frac{i}{4} \bar\psi\Gamma^{MNP} \Gamma^Q \Gamma^A \Gamma^B \eps T^A_{MNP} D_Q \phi^B\cr
(A7c) &=& - i \bar\psi \Gamma^{MNP}\Gamma^A\Gamma^B\eta T^A_{MNP}\phi^B
\eea
We now find that the following terms cancel,
\bea
0 &=& (A1) + (A2e)\cr
0 &=& (A2c) + (A2h)
\eea
Now we will expand out $(A2i)$ by using 
\bea
5 \Gamma^M D_M \eta &=& - \frac{R}{4} \eps + \frac{1}{8} \Gamma^{MN} \Gamma^{AB} \eps W_{MN}^{AB} - \frac{3}{4} \Gamma^A \Gamma^{MNP} \eta T^A_{MNP} - \frac{1}{8} \Gamma^A \Gamma^{UVW} \Gamma^M \eps D_M T^A_{UVW}
\eea
that is a direct consequence of (\ref{CKE}) as we show in the appendix. Here $W_{MN}^{AB}$ is a field strength of the R-gauge field as defined in (\ref{W}). Then we get
\bea
(A2ia) &=& - \frac{R}{5} i \bar\eps\Gamma^A \psi\cr
(A2ib) &=& \frac{i}{10} \bar\psi \Gamma^A \Gamma^{BC} \Gamma^{MN} \eps W_{MN}^{BC} \phi^A\cr
(A2ic) &=& - \frac{3i}{5} \bar\psi \Gamma^A \Gamma^B \Gamma^{RST} \eta T^B_{RST} \phi^A\cr
(A2id) &=& - \frac{i}{10} \bar\psi \Gamma^A \Gamma^B \Gamma^{RST} \Gamma^M \eps (D_M T^B_{RST}) \phi^A
\eea
Now we collect terms as follows,
\bea
(A2g) + (A2ib) &=& \frac{i}{2} \bar\psi \Gamma^{MN} \(\Gamma^{E} \delta^{FG} + \frac{1}{5} \Gamma^{G} \Gamma^{EF}\) \eps W_{MN}^{EF} \phi^G\cr
&=& - \frac{i}{2} \bar\psi \Gamma^{MN} \(\delta^{GE} - \frac{1}{5} \Gamma^G \Gamma^E\) \Gamma^F \eps W^{EF}_{MN} \phi^G\cr
(A7c) + (A2ic) &=& - i \bar\psi \Gamma^{MNP} \(\Gamma^A \Gamma^B + \frac{3}{5} \Gamma^B\Gamma^A\)\eta T^A_{MNP} \phi^B\cr
&=& - 2 i \bar\psi \Gamma^{MNP} \(\delta^{AB} - \frac{1}{5} \Gamma^A\Gamma^B\) \eta T^B_{MNP} \phi^A
\eea
We also get
\bea
(A2d) + (A7b) &=& - \frac{i}{4} \bar\psi \Gamma^{RST}\Gamma^M \Gamma^A \Gamma^B \eps T^B_{RST} D_M \phi^A\cr
&& - \frac{i}{4} \bar\psi\Gamma^{MNP} \Gamma^Q \Gamma^A \Gamma^B \eps T^A_{MNP} D_Q \phi^B\cr
&=& - \frac{i}{2} \bar\psi \Gamma^{RST} \Gamma^M \eps T^A_{RST} D_M \phi^A
\eea
We conclude that the contribution to the supersymmetry variation of the matter fields Lagrangian that comes from the abelian terms is given by\footnote{I thank Dongsu Bak for that he carried out a similar computation to this one for the abelian M5 brane in an unfinished separate project several years ago.} 
\bea
\delta \L_m &=& \bar\psi \(\delta^{AB} - \frac{1}{5} \Gamma^A\Gamma^B\) \chi^B \phi^A - \frac{iR}{5} \bar\psi \Gamma^A \eps \phi^A + \mu^{AB} i \bar\psi\Gamma^A\eps\phi^B
\eea
where
\bea
\chi^A &=& - \frac{i}{2} \Gamma^{MN} \Gamma^B \eps W^{AB}_{MN} - 2 i \Gamma^{MNP} \eta T^A_{MNP} + \frac{i}{2} \Gamma^{UVW} \Gamma^M \eps D_M T^A_{UVW}
\eea
For this variation to vanish we shall take 
\ben
\mu^{AB} &=& \frac{R}{5} \delta^{AB} - D^{AB}\label{mu}
\een
where $D^{AB}$ is a symmetric tensor that satisfies
\bea
\chi^A - \frac{1}{5} \Gamma^A \Gamma^B \chi^B &=& \Gamma^B \eps D^{AB}
\eea
We can also see that $D^{AB}$ shall be traceless by contracting both sides with $\Gamma^A$ from the left. 

To better understand the variation of the matter field Lagrangian, we will now study the following term in $\delta \L_b$ Lagrangian
\bea
\L_{T} &=& H^{+MNP} T^A_{MNP} \phi^A
\eea
Its has the following supersymmetry variation
\bea
\delta \L_{T} &=& - 3 i \bar\eps\Gamma^{NP} \psi D^M\(T^A_{MNP} \phi^A\) + i \bar\eps\Gamma^A\psi H^{+MNP} T^A_{MNP}
\eea
We are now interested in the first term that we expand out in two terms 
\bea
- 3 i \bar\eps\Gamma^{NP} \psi \(D^M T^A_{MNP}\) \phi^A\cr
- 3 i \bar\eps\Gamma^{NP} \psi T^A_{MNP} D^M\phi^A
\eea
The second term cancels $(A2d)+(A7b)$ by using the fact that $T^A_{MNP}$ is antiselfdual and the first term combines with $(A2id)$ to give the last term in $\chi^A$ as
\bea
\chi^A_{last} &=& 3 i \Gamma^{MN} \eps D^P T^A_{MNP}
\eea
Then finally the term
\bea
(A8) &=& i \bar\eps\Gamma^A\psi H^{+MNP} T^A_{MNP}
\eea
combines with other terms into a cancelation,
\bea
(A2a)+(A7a)+(A8) &=& 0
\eea
which uses the gamma matrix identity\footnote{We discovered this gamma matrix identity by using GAMMA \cite{Gran:2001yh}.}
\bea
- \frac{1}{8\cdot 12} \Gamma^M \Gamma^{RST} \Gamma^{UVW} \Gamma_M - \frac{1}{4\cdot 12} \Gamma^{UVW} \Gamma^{RST} &=&  \frac{1}{2} \(g^{RST,UVW} + \frac{1}{6} \Gamma^{RSTUVW}\)
\eea
Then we notice that $\Gamma^{RSTUVW} = \eps^{RSTUVW} \Gamma$ and that $\Gamma \eps = - \eps$ so that when this acts on $\eps$ it will generate a projection onto the selfdual part of $T^A_{UVW}$ which is zero.

We are left with $(A2b)$ and we have added one term that we need to subtract again. Combining this with the commutator term obtained previosuly, we are now ready to write down our final result for the variation of $\L_m$. It is given by
\bea
\delta \L_m &=& - \frac{1}{2} \delta B^{NP} D^M H^+_{MNP} - \delta\(H^{+MNP} T^A_{MNP} \phi^A\)
\eea

\section{The supersymmetry variation of $\L_b$}
Let us begin by making a supersymmetry variation of $\t\L_b$ given by 
\bea
\t\L_b &=& \frac{1}{24} h_{MNP}^2 + \frac{1}{6} H^{+MNP} \(h^-_{MNP} + w^-_{MNP} + 6 T^A_{MNP}\phi^A\)\cr
&& + \frac{1}{6} h^{-MNP} w^+_{MNP} + \frac{1}{24} w_{MNP}^2\cr
&& + \lambda^{-MNP} \(H^+_{MNP} + h^+_{MNP} + w^+_{MNP}\)\cr
&& + \frac{1}{48} \eps^{MNPQRS} F_{MN} W_{PQRS}
\eea
where we omit the Chern-Simons term. Here 
\bea
\delta H^+_{MNP} &=& - \delta h^+_{MNP} - \delta w^+_{MNP}
\eea
We get
\bea
\delta \t\L_b &=& \frac{1}{12} h^{+MNP} \delta h_{MNP} + \frac{1}{12} h^{-MNP} \delta h_{MNP}\cr
&& - \frac{1}{6} \(h^{-MNP} + w^{-MNP}\) \(\delta h_{MNP} + \delta w_{MNP}\) \cr
&& + \frac{1}{6} H^{+MNP} \(\delta h_{MNP} + \delta w_{MNP}\)\cr
&& + \frac{1}{12} w^{+MNP} \delta w_{MNP} + \frac{1}{12} w^{-MNP} \delta w_{MNP}\cr
&& + \delta \(H^{+MNP} T^A_{MNP} \phi^A\)\cr
&& + \frac{1}{6} h^{-MNP} \delta w_{MNP} + \frac{1}{6} w^{+MNP} \delta h_{MNP}\cr
&& + \frac{1}{48} \eps^{MNPQRS} \delta \(F_{MN} W_{PQRS}\)
\eea
The coefficients of selfdual and antiselfdual components now conspire so that we obtain several terms that are wedge products between three-forms,
\ben
\delta \t\L_b &=& \delta \(H^{+MNP} T^A_{MNP} \phi^A\)\cr
&& + \frac{1}{12\cdot 6} \eps^{MNPRST} h_{RST} \delta h_{MNP}\cr
&& + \frac{1}{12\cdot 6} \eps^{MNPRST} w_{RST} \delta w_{MNP}\cr
&& + \frac{1}{6} H^{+MNP} \delta \(h_{MNP} + w_{MNP}\)\cr
&& + \frac{1}{12\cdot 6} \eps^{MNPRST} w_{RST} \delta h_{MNP}\cr
&& + \frac{1}{48} \eps^{MNPQRS} \delta \(F_{MN} W_{PQRS}\)\label{third}
\een
We now expand out the term in the second line
\bea
\delta \L_2 &=& \frac{1}{12\cdot 6} \eps^{MNPRST} h_{RST} \delta h_{MNP}\cr &=& \frac{1}{24} \eps^{MNPRST} D_M h_{RST} \delta B_{NP}
\eea
Now we use (\ref{bhid}) and we get
\bea
\delta \L_2 &=& - \frac{1}{48} \eps^{MNPRST} F_{MR} w_{ST} \delta B_{NP}
\eea
To proceed we want neither $F_{MN}$ nor $w_{MN}$ to have any component in the $v_M$ direction. This is solved for $F_{MN}$ by imposing the gauge fixing condition $v^M A_M = 0$ and by demanding $\L_v A_M = 0$ since this implies that $F_{MN} v^N$. For $w_{MN}$ we need to assume that $v_M v^M$ is constant, which implies that our six-manifold shall be a K-contact manifold, since only then do we also get $w_{MN} v^N = 0$. This is easy to see. First we note that $\L_v v_N = v^M \nabla_M v_N + (\nabla_N v_M) v^M$ and then we use the Killing equation $\nabla_M v_N + \nabla_N v_M = 0$ on the second term, and we see that it cancels the first term so $\L_v v_N =0$. Next we note that $v^M w_{MN} = v^M \nabla_M v_N - v^M \nabla_N v_M = \L_v v_N - \nabla_N (v^M v_M)$ and this vanishes only if $v^M v_M$ is constant. As now no component in the direction of $v_M$ comes from neither $F_{MR}$ nor from $w_{ST}$ it must come from $\delta B_{NP}$. So we can replace $\delta B_{NP} \rightarrow Q_P^S \delta B_{NS} = \delta A_N v_P/v^2$, 
\bea
\delta \L_2 &=& - \frac{1}{48 v^2} \eps^{MNPRST} F_{MR} w_{ST} v_P \delta A_N 
\eea
This variation is now precisely canceled by the variation of the Chern-Simons term
\bea
\L_{CS} &=& - \frac{1}{24 v^2} \eps^{MNPRST} \tr\(A_M \nabla_N A_P - \frac{2i e}{3} A_M A_N A_P\) w_{RS} v_T
\eea
so that we have
\bea
\delta \L_{CS} + \delta \L_2 &=& 0
\eea

The term in the third line in (\ref{third}) is worrisome as it can not be canceled by any other term. Fortunately it is identically zero as the following detailed computation shows,
\bea
\eps^{MNPRST} w_{RST} \delta w_{MNP} &=& \eps^{MNPRST} W_{RSTU} \delta W_{MNPV} v^U v^V\cr
&=& - e \eps^{MNPRST} W_{RSTU} \bar\eps \Gamma_{MNPV} [\psi,\phi^A] v^U v^V\cr
&=& - 18 e W_{RSTU} \bar\eps \Gamma^{RS} [\psi,\phi^A] v^U v^T\cr
&=& 0
\eea
where we have used the gamma matrix identity
\bea
\Gamma_{MNPV} \Gamma^{MNPRST} &=& 18 \Gamma^{[RS} \delta^{T]}_V
\eea
After all these considerations, our result collapses to 
\bea
\delta \L_b &=& \delta \(H^{+MNP} T^A_{MNP} \phi^A\)\cr
&& + \frac{1}{6} H^{+MNP} \delta \(h_{MNP} + w_{MNP}\)\cr
&& + \frac{1}{12\cdot 6} \eps^{MNPRST} w_{RST} \delta h_{MNP} + \frac{1}{24} \eps^{MNPRST} D_M \(\delta B_{NQ} v^Q\) W_{PRST}\cr
&& + \frac{1}{48} \eps^{MNPQRS} F_{MN} \delta W_{PQRS}
\eea
The two terms on the third line cancel up to a Lie derivative,\footnote{The gamma matrix relations that are used here are 
\bea
\eps^{MNPQRS} \Gamma_{RS} &=& 2 \Gamma^{MNPQ} \Gamma\cr
5 \Gamma^{[PQRS} v^{M]} &=& \Gamma^{PQRS} v^M - 4 \Gamma^{[PQR|M|} v^{S]}
\eea} 
\bea
&& \frac{1}{36} \eps^{MNPRST} \delta h_{RST} w_{MNP} + \frac{1}{24} \eps^{MNPRST} D_M \(\delta B_{NU} v^U\) W_{PRST}\cr
&=& \frac{i}{24} \bar\eps\Gamma^{MNPQ} \psi \(v^S D_S W_{MNPQ} + 4 D_Q\(W_{MNPS} v^S\) - 4 v^S D_Q W_{MNPS}\)\cr
&=& \frac{i}{24} \bar\eps\Gamma^{MNPQ} \psi \L_v W_{MNPQ} 
\eea
Putting this Lie derivative to zero as a constraint that we impose on top of the Lagrangian, we can now write the variation of the Lagrangian as 
\bea
\delta \L_b &=& \delta \(H^{+MNP} T^A_{MNP} \phi^A\)\cr
&& + \frac{1}{6} H^{+MNP} \delta h_{MNP}\cr
&& - \frac{1}{36} \eps^{MNPRST} H^+_{MNP} \delta W_{RSTQ} v^Q\cr
&& + \frac{1}{48} \eps^{MNPRST} F_{MN} \delta W_{PRST}
\eea
We will now argue that the two last terms cancel upon using the constraint
\ben
F_{MN} &=& \(H^+_{MNP} + 6 T^A_{MNP} \phi^A\) v^P\label{Co}
\een
To this end we start by making the following observation that if we define selfdual parts of $W_{MNPQ}$ as
\bea
W^{\pm}_{MNPQ} &=& \frac{1}{2} \(W_{MNPQ} \pm \frac{1}{6} \eps_{[MNP}{}^{RST} W_{|RST|Q]}\)
\eea
then we can write the last term in the Lagrangian in the following form
\bea
\frac{1}{48} \eps^{MNPQRS} F_{MN} W_{PQRS} &=& \frac{1}{24} \eps^{MNPQRS} F_{MN} W^-_{PQRS}
\eea
This is a consequence of $W_{MNP}{}^P = 0$ that follows if one assumes that $W_{MNPQ}$ is totally antisymmetric in all four indices. Now we use the constraint (\ref{Co}) and then this term becomes proportional to
\bea
\(H^{+PQR} - 6 T_A^{-PQR}\phi^A\) v^S \delta W^-_{PQRS} &=& H^{+PQR} v^S \delta W^-_{PQRS}
\eea
so the upshot is that by using (\ref{Co}) we have
\bea
\frac{1}{48} \eps^{MNPQRS} F_{MN} \delta W_{PQRS} &=& \frac{1}{48} \eps^{MNPQRS} H^+_{MNU} v^U \delta W_{PQRS}
\eea
and now it is easy to see that this cancels against the term
\bea
- \frac{1}{36} \eps^{MNPRST} H^+_{MNP} \delta W_{RSTQ} v^Q
\eea
by noting the following identity
\bea
3!4! H^+_{[MNQ} \delta W_{RSTP]} &=& 2!4! H^+_{[MN|Q|} \delta W_{RSTP]} - 3!3! H^+_{[MNP} \delta W_{RST]Q}
\eea
and the fact that the left-hand side is identically zero because we antisymmetrize in seven indices, each of which takes six different values. So we are left with the variation
\bea
\delta \L_b &=& \delta\(H^{+MNP} T^A_{MNP} \phi^A\) \cr
&& + \frac{1}{2} \delta B^{NP} D^M H^+_{MNP}
\eea
so that this cancels the variation of $\L_m$,
\bea
\delta \L_b + \delta \L_m &=& 0
\eea

\section{Equations of motion} 
We will derive the on-shell Bianchi identity for $H^+_{MNP}$ that is required for on-shell closure of the supersymmetry variations when we act twice with supersymmetry variations on $H^+_{MNP}$. We will show that it arises as an equation of motion that we derive from the Lagrangian $\L = \L_b + \L_m$. This is thus a consistency check.

\subsection*{The equation of motion for $A_M$}
Varying $A_M$ we get
\bea
0 &=& \frac{1}{24} \eps^{MNPQRS} \(D_N W_{PQRS} - \frac{1}{v^2} F_{[NQ} w_{RS}v_{P]}\)\cr
&& - i e \[D^M \phi^A,\phi^A\] - \frac{e}{2} \{\bar\psi,\Gamma^M \psi\}
\eea
Let us dualize the equation of motion,
\ben
5 \(D_{[M} W_{NPQR]} - \frac{1}{v^2} F_{[MP} w_{QR} v_{N]}\) &=& \eps_{MNPQRS} \(i e [D^S \phi^A,\phi^A] + \frac{e}{2} \{\bar\psi,\Gamma^S \psi\}\)\label{weqom}
\een
Contracting with $v^R$ we get
\ben
&& 4 \(D_{[M} w_{NPQ]} - \frac{1}{4} F_{[MN} w_{PQ]}\) - \L_v W_{MNPQ} \cr
&&= i e \eps_{MNPQRS} [D^S \phi^A,\phi^A] v^R + \frac{e}{2} \eps_{MNPQRS} \{\bar\psi,\Gamma^S\psi\} v^R\label{w}
\een

\subsection*{The equation of motion for $b_{MN}$}
Varying $h_{MNP}$ according to our postulated rule, $\delta h_{MNP} = 3 D_{[M} \delta b_{NP]}$, we get
\ben
D_M \(h^{MNP} + 2 H^{+MNP} + 2 w^{+MNP} + 12 \lambda^{-MNP}\) &=& \frac{1}{2} \eps^{MNPRST} F_{RS} w_{TM}\label{bMN}
\een

\subsection*{The equation of motion for $H^{+MNP}$}
Varying $H^{+MNP}$ we get the selfduality equation of motion
\ben
h^-_{MNP} + w^-_{MNP} + 6 T^A_{MNP} \phi^A + 6 \lambda^-_{MNP} &=& 0\label{SelfDual}
\een

\subsection*{The equation of motion for $\lambda^{-MNP}$}
Varying $\lambda^{-MNP}$ we get a constraint that relates $H^+$ to $h^+ + w^+$,
\bea
H^+_{MNP} &=& - h^+_{MNP} - w^+_{MNP}
\eea
This constraint is supersymmetry invariant by itself.

\subsection*{The equation of motion for $W_{MNPQ}$}
Varying $W_{MNPQ}$ we get
\bea
H^+_{UVT} v^T - F_{UV} - h^-_{UVT} v^T + \frac{1}{12} \eps_{UVTMNP} w^{MNP} v^T - 6 \lambda^-_{UVT} v^T &=& 0
\eea
Then if we use (\ref{SelfDual}), then this equation reduces to
\ben
F_{UV} &=& \(H^+_{UVT} + 6 T^A_{UVT}\phi^A\) v^T\label{FH}
\een
To see this, we need to establish that the remaining terms cancel. Namely we need to establish that 
\bea
w^-_{UVT} v^T + \frac{1}{12} \eps_{UVTMNP} w^{MNP} v^T &=& 0
\eea
but this is an identity that collapses to $w_{UVT} v^T = W_{UVTR} v^T v^R = 0$ by using the defintion
\bea
w^-_{UVT} &=& \frac{1}{2} \(w_{UVT} - \frac{1}{6} \eps_{UVTMNP} w^{MNP}\)
\eea
of the antiselfdual part.

We now notice that $h+w-6T^A\phi^A$ is selfdual, which means that  
\bea
h+w-6T^A\phi^A &=& h^+ + w^+
\eea
because $T^A$ is antiselfdual. Then we can use this in the constraint $H^+ = - h^+ - w^+$ to get
\bea
H^+ + 6 T^A \phi^A &=& - h - w
\eea
which means that we can express (\ref{FH}) as 
\ben
F_{MN} &=& - h_{MNP} v^P\label{Fh}
\een
where we have used that $w_{MNP} v^P = W_{MNPQ} v^P v^Q = 0$ for the nonchiral $w_{MNP}$. The equation (\ref{Fh}) is invariant under supersymmetry variations up to a Lie derivative that we constrain to be zero. Namely the variation of the right-hand side is
\bea
- \delta h_{MNP} v^P &=& 3 (D_{[M} \delta B_{NP]}) v^P \cr
&=& 2 D_{[M} \(\delta B_{N]P} v^P\) + v^P D_P \delta B_{MN} - 2 D_{[M} v^P \delta B_{N]P}\cr
&=& \delta F_{MN} + \L_v \delta B_{MN}
\eea

\subsection{The on-shell Bianchi identity}
If we eliminate $\lambda^-_{MNP}$ from (\ref{SelfDual}) and insert that into (\ref{bMN}) then we get
\bea
D_M \(h^{+MNP} - h^{-MNP} + 2 H^{+MNP} + 2(w^{+MNP} - w^{-MNP}) - 12 T_A^{-MNP} \phi^A\) &=& \frac{1}{4} \eps^{MNPRST} F_{RS} w_{TM}
\eea
where we may now notice how the coefficients in the Lagrangian conspire so that this becomes nonchiral three-forms once we dualize the three-form expression in the parentesis. We then get 
\bea
D_{[M} \(h_{RST]} + 2 H^+_{RST]} + 2 w_{RST]} + 12 T^A_{RST]} \phi^A\) &=& \frac{1}{4} F_{[RS} w_{TM]}
\eea
Now we use the Bianchi identity $D_{[M} h_{RST]} = 0$ and we get
\bea
D_{[M} \(H^+_{RST]} + 6 T^A_{MNP]} \phi^A\) &=& \frac{1}{4} F_{[RS} w_{TM]} - D_{[M} w_{RST]}
\eea
and finally we use the equation of motion for $A_M$ obtained in (\ref{w}) and we arrive at the on-shell Bianchi identity
\bea
D_{[M} \(H^+_{NPQ]} + 6 T^A_{NPQ]} \phi^A\) &=& - \frac{i e}{4} \eps_{MNPQRS} [D^S \phi^A,\phi^A] v^R - \frac{e}{8} \eps_{MNPQRS} \{\bar\psi,\Gamma^S\psi\} v^R
\eea
that is the equation of motion that is required in order to close the supersymmetry variations on $H^+_{MNP}$ as was originally shown in \cite{Lambert:2010wm}, but here this equation of motion was derived from the Lagrangian. 

Our computation is the same in spirit as that in \cite{Lambert:2019diy}, but it differs in the details. In \cite{Lambert:2019diy} in place of our $h_{MNP}$ there appears instead expressions directly in terms of a nonabelian two-form $b_{MN}$ (using our notation). This is of course more attractive than our approach since it makes the equations explicit. However, their nonabelian two-form gauge potential appears in places where we would not expect that a gauge potential would appear explicitly, in the Lagrangian and in the supersymmetry variation of $W_{MNPQ}$. Those quantities shall transform gauge covariantly, which is why we have chosen to set up the things in a different way from \cite{Lambert:2019diy}.

Since unlike \cite{Lambert:2019diy} we have allowed $v_M$ to have a nonvanishing derivative, reflected in having a nonvanishing two-form $w_{MN}$, this has led us to discover a new Chern-Simons term $A \wedge F \wedge v \wedge w$. In \cite{Cordova:2013cea} it was shown that if one puts M5 brane on $S^3 \times M_3$ for some euclidean three-manifold $M_3$ and if one performs dimensional reduction on $S^3$ (possibly a squashed $S^3$, which would be reflected in having a nontrivial rescaling between our $w_{MN}$ and $v_M$), one gets a complex Chern-Simons theory on $M_3$. It seems plausible that our real Chern-Simons term could be somehow related to this complex Chern-Simons theory on euclidean $M_3$. In our computation we have assumed Lorentzian signature, so it seems like we would not get a complex Chern-Simons in our Lorentzian computation. This needs to be studied further.

\section{The relation with the nonchiral Lagrangian}
If we integrate out $\lambda^-_{MNP}$, then that will amount to replacing $H^+_{MNP}$ with $- h^+_{MNP} - w^+_{MNP}$ in the Lagrangian. If we do that, then we can recast the Lagrangian in the form
\bea
\L_b &=& - \frac{1}{24} g_{MNP}^2 - \frac{1}{12} \cdot \frac{1}{6} \eps^{MNPRST} g_{MNP} C_{RST}\cr
&& + \frac{1}{2} w^{MNP} T^A_{MNP} \phi^A\cr
&& + \frac{1}{48} \eps^{MNPQRS} F_{MN} W_{PQRS}
\eea
where $C_{MNP} = w_{MNP} + 6 T^A_{MNP} \phi^A$. 

If we truncate to the sector $W_{MNPQ} = 0$ by hand in this Lagrangian, then we recover the traditional nonchiral Lagrangian \cite{Witten:1996hc}
\bea
\t\L_b &=& - \frac{1}{24} F_{MNP}^2 + \frac{1}{12} \cdot \frac{1}{6} \eps^{MNPRST} F_{MNP} C_{RST}
\eea
where $F_{MNP} = - h_{MNP} + C_{MNP}$ and where only the selfdual part of $h_{MNP}$ is coupled to the three-form field 
\bea
C_{MNP} &=& - 6 T^A_{MNP} \phi^A
\eea
Putting $W_{MNPQ} = 0$ is not a consistent truncation in the nonabelian case because the supersymmetry variation of $W_{MNPQ}$ is nonzero. But it is a consistent truncation in the abelian case and there this nonchiral action $\int \(\t\L_b + \L_m\)$ is fully supersymmetric once we replace $\delta \psi = \frac{1}{12} \Gamma^{MNP} \eps H^+_{MNP} + ...$ with $\delta \psi = \frac{1}{12} \Gamma^{MNP} \eps H_{MNP} + ...$ where $H_{MNP} := - h_{MNP}$. 

The role of $W_{MNPQ}$ is to promote the constraint (\ref{Fh}) to an equation of motion, which has the advantage that we can derive the equations of motion by varying the fields $A_M$ and $h_{MNP}$ as independent fields in the Lagrangian. 

\section{Closure of supersymmetry variations}
Here we assume that the supersymmetry parameter is commuting and compute $\delta^2$ on each field. Since we have introduced many auxiliary fields with no accompanying fermionic auxiliary fields, we do not necessarily expect closure on all these auxiliary fields. 

\subsection*{Closure on $\phi^A$}
\bea
\delta^2 \phi^A &=& - i S^M D_M \phi^A - 4 i \bar\eps\eta \phi^A - 4 i \bar\eps\Gamma^{AB}\eta \phi^B - i e [\phi^A,\Lambda]
\eea
where the gauge parameter is 
\bea
\Lambda &=& - i \bar\eps \Gamma^M \Gamma^A \eps \phi^A v_M
\eea

\subsection*{Closure on $A_M$}
\bea
\delta^2 A_M &=& - i S^T \(H^+_{MNT} + 6 T^A_{MNT} \phi^A\) v^N + D_M \Lambda\cr
&& - 2 i \bar\eps\Gamma^A \Gamma_M \(\L_v \eps\) \phi^A
\eea
We have closure up to a gauge transformation if we impose the constraint 
\ben
F_{MN} &=&  \(H^+_{MNP}+6T^A_{MNP}\phi^A\) v^P\label{fmn}
\een
This constraint is consistent with what we found in (\ref{FH}) and in (\ref{Co}), so now we have found this constraint by three different computations, thereby making it rather convincing that it must be correct.

\subsection*{Closure on $W_{MNPQ}$?}
Using the equation of motion (\ref{weqom}) we get
\bea
\delta^2 W_{MNPQ} &=& 5 i S^R D_{[M} W_{NPQR]}\cr
&=& - i S^R D_{R} W_{NPQM} + 4 i S^R D_{[M} W_{NPQ]R}\cr
&=& - i S^R D_R W_{NPQM} - 4 i (D_{[M} S^R) W_{NPQ]R}\cr
&& + 4 i D_{[M} \(W_{NPQ]R} S^R\)
\eea
which we can write as
\bea
\delta^2 W_{MNPQ} &=& - i \L^A_S W_{MNPQ} + D_{[M} \lambda_{NPQ]}
\eea
where 
\bea
\lambda_{NPQ} &=& 4 i W_{NPQR} S^R
\eea
and $\L_S^A$ is a Lie derivative where gauge covariant derivatives are used. We were unable to show that the second term is a gauge symmetry of the Lagrangian. However, we may eliminate this problem by simply integrating out $W_{MNPQ}$ that will impose the constraint (\ref{fmn}).

\section{Deriving the fermionic equation of motion from selfduality} 
We would like to show that we get the fermionic equation of motion by making a supersymmetry variation of the selfduality equation of motion
\ben
\(h_{MNP} + w_{MNP} + 6 T^A_{MNP} \phi^A\)^- &=& 0\label{SelfDuality}
\een
if we vary $h_{MNP}$ according to the rule $\delta h_{MNP} = 3 D_{[M} \delta b_{NP]}$. We use
\bea
\delta h_{MNP}^- &=& - \frac{i}{2} D_Q \(\bar\eps \Gamma_{MNP} \Gamma^Q \psi\)\cr
&=& - \frac{i}{2} \bar\eps \Gamma_{MNP} \Gamma^Q D_Q \psi + \frac{i}{16} \bar\eps \Gamma_Q \Gamma^{RST} \Gamma^A \psi T^A_{RST}\cr
\delta w_{MNP}^- &=& - \frac{e}{2} \bar\eps \Gamma_{MNP} \Gamma^Q [\psi,\phi^A] v_Q
\eea
Making a supersymmetry variation of (\ref{SelfDuality}), we then get
\bea
- i \Gamma_{MNP} \Gamma^Q D_Q \psi + \frac{i}{8} \Gamma_Q \Gamma^{RST} \Gamma_{MNP}\Gamma^Q \Gamma^A \psi T^A_{RST} - e \Gamma_{MNP} \Gamma^Q [\psi,\phi^A] v_Q + 12 i T^A_{MNP} \Gamma^A \psi &=& 0
\eea
Now contracting from the left with $\Gamma^{MNP}$ and using 
\bea
\Gamma^{MNP} \Gamma_{MNP} &=& - 120\cr
\Gamma^{MNP} \Gamma_Q \Gamma^{RST} \Gamma_{MNP} \Gamma^Q &=& 144 \Gamma^{RST}
\eea
we get
\bea
i 120 \Gamma^Q D_Q \psi + 18 i \Gamma^{RST}\Gamma^A\psi T^A_{RST} + 120 e \Gamma^Q[\psi,\phi^A]v_Q + 12 i \Gamma^{MNP}\Gamma^A \psi T^A_{MNP} &=& 0
\eea
Then 
\bea
i \Gamma^Q D_Q \psi + \frac{i}{4} \Gamma^{MNP} \Gamma^A \psi T^A_{MNP} + e \Gamma^Q [\psi,\phi^A] v_Q &=& 0
\eea
which agrees with the fermionic equation of motion that we obtain from the matter fields Lagrangian.

We have assumed that the infinitesimal variation of $h_{MNP}$ is on the form $\delta h_{MNP} = 3 D_{[M} \delta b_{NP]}$ and we have seen that this assumption takes the selfduality equation of motion to the expected fermionic equation of motion. This thus seems like the correct assumption for the infinitesimal variation of $h_{MNP}$.

\section*{Acknowledgements}
I would like to thank Neil Lambert for explaining his ideas in his recent work to me, and Ulf Gran for assistance on how to use his computer program GAMMA \cite{Gran:2001yh}. This work was supported in part by NRF Grant 2020R1A2B5B01001473 and NRF Grant 2020R1I1A1A01052462.

\appendix
\section{Derivation of the Fierz identity}
For the M5 brane we have the following Fierz identity for two anticommuting fermions $\psi^a$ and $\psi^b$ where $a,b, ...$ are adjoint gauge group indices,
\bea
\psi^a \bar\psi^b &=& \(- \frac{1}{16} \bar\psi^b \Gamma^M \psi^a \Gamma_M + \frac{1}{16} \bar\psi^b \Gamma^M \Gamma^A \psi^a \Gamma_M \Gamma^A + \frac{1}{192} \bar\psi^b \Gamma^{MNP} \Gamma^{AB} \psi^a \Gamma_{MNP} \Gamma^{AB}\) P_-
\eea
where $P_- = \frac{1}{2} (1-\Gamma)$. Then we get
\bea
\Gamma_{PQ} \psi^a \bar\psi^b \Gamma^Q \psi^c &=& \frac{3}{16} (\bar\psi^b \Gamma^M\psi^a) \Gamma_{PM} \psi^c + \frac{5}{16} (\bar\psi^b \Gamma_P \psi^a) \psi^c\cr
&& + \frac{3}{16} (\bar\psi^b \Gamma^M\Gamma^A \psi^a) \Gamma_{PM} \Gamma^A \psi^c + \frac{5}{16} (\bar\psi^b \Gamma_P \Gamma^A \psi^a) \Gamma^A \psi^c\cr
&& + \frac{1}{192} (\bar\psi^b \Gamma_{RST} \Gamma^{AB} \psi^a) \Gamma^{RST} \Gamma_P \Gamma^{AB} \psi^c\cr
\Gamma^A \psi^a \bar\psi^b \Gamma_P \Gamma^A \psi^c &=& \frac{5}{16} (\bar\psi^b \Gamma^M \psi^a) \Gamma_{PM} \psi^c - \frac{5}{16} (\bar\psi^b\Gamma_P \psi^a) \psi^c\cr
&& - \frac{3}{16} (\bar\psi^b\Gamma^M\Gamma^A\psi^a) \Gamma_{PM} \Gamma^A \psi^c + \frac{3}{16}(\bar\psi^b\Gamma_P\Gamma^A\psi^a)\Gamma^A\psi^c\cr
&& - \frac{1}{192} (\bar\psi^b \Gamma_{RST} \Gamma^{AB} \psi^a) \Gamma^{RST} \Gamma_P \Gamma^{AB} \psi^c
\eea
Adding these, we get the following identity
\bea
\Gamma^{PQ} \psi^a (\bar\psi^b\Gamma_Q \psi^c) + \Gamma^A \psi^a (\bar\psi^b\Gamma^P\Gamma^A\psi^c) &=& \frac{1}{2} \(\Gamma^{PQ} \psi^c (\bar\psi^b\Gamma_Q \psi^a) + \Gamma^A \psi^c (\bar\psi^b\Gamma^P\Gamma^A\psi^a)\)
\eea
So when we contract the gauge indices $a,b,c$ with totally antisymmetric structure constants $f_{abc}$ of the gauge group, we get
\bea
\(\Gamma^{PQ} \psi^a (\bar\psi^b\Gamma_Q \psi^c) + \Gamma^A \psi^a (\bar\psi^b\Gamma^P\Gamma^A\psi^c)\) f_{abc} &=& 0
\eea
and this is precisely the identity we need for the cubic terms in the supersymmetry variation of the Lagrangian to vanish.

\section{A consequence of the Killing spinor equation}
From (\ref{CKE}) we have
\bea
6 \Gamma^M D_M \eta = \Gamma^M \Gamma^N D_M D_N \eps = D_M^2 \eps + \frac{R}{4} \eps - \frac{1}{8} W_{MN}^{AB} \Gamma^{MN} \Gamma^{AB} \eps
\eea
and
\bea
D_M^2 \eps = \Gamma^M D_M \eta - \frac{3}{4} \Gamma^A \Gamma^{RST} \eta T^A_{RST} - \frac{1}{8} \Gamma^A \Gamma^{PQR} \Gamma^M \eps D_M T^A_{PQR}
\eea
where
\ben
W^{AB}_{MN} &=& \partial_M V_N^{AB} - \partial_N V_M^{AB}\label{W}
\een
is the field strength of the R-gauge field background potential $V_M^{AB}$. By taking these two results together, we get
\bea
5 \Gamma^M D_M \eta &=& - \frac{R}{4} \eps + \frac{1}{8} \Gamma^{MN} \Gamma^{AB} \eps W_{MN}^{AB} - \frac{3}{4} \Gamma^A \Gamma^{MNP} \eta T^A_{MNP} - \frac{1}{8} \Gamma^A \Gamma^{UVW} \Gamma^M \eps D_M T^A_{UVW}
\eea

\end{document}